\documentclass[fleqn,usenatbib]{mnras}

\usepackage{graphicx}    
\usepackage{amsmath}     
\usepackage{amssymb}     
\usepackage{newtxtext,newtxmath} 
\usepackage{hyperref}
\usepackage{orcidlink}
\usepackage{natbib}

\title[Model Averaging in Pulsar Timing]{Use Model Averaging instead of Model Selection in Pulsar Timing}

\author[Rutger van Haasteren\orcidlink{0000-0002-6428-2620}]{
    Rutger van Haasteren\orcidlink{0000-0002-6428-2620}\thanks{E-mail: rutger@vhaasteren.com}\\
    Max-Planck-Institut f{\"u}r Gravitationsphysik (Albert-Einstein-Institut), Callinstra{\ss}e 38, D-30167, Hannover, Germany\\
    Leibniz Universit{\"a}t Hannover, D-30167, Hannover, Germany
}

\date{Accepted 2024 October 31. Received 2024 October 15; in original form 2024 September 16}
\pubyear{2024}

\begin{document}

\maketitle

\begin{abstract}
    Over the past decade and a half, adoption of Bayesian inference in pulsar timing analysis has led to increasingly sophisticated models. The recent announcement of evidence for a stochastic background of gravitational waves by various pulsar timing array projects highlighted Bayesian inference as a central tool for parameter estimation and model selection. Despite its success, Bayesian inference is occasionally misused in the pulsar timing community.
    A common workflow is that the data is analyzed in multiple steps: a first analysis of single pulsars individually, and a subsequent analysis of the whole array of pulsars. A mistake that is then sometimes introduced stems from using the posterior distribution to craft the prior for the analysis of the same data in a second step, a practice referred to in the statistics literature as ``circular analysis.'' This is done to prune the model for computational efficiency.
    Multiple recent high-profile searches for gravitational waves by pulsar timing array (PTA) projects have this workflow.
    This letter highlights this error and suggests that Spike and Slab priors can be used to carry out model averaging instead of model selection in a single pass. Spike and Slab priors are proved to be equal to Log-Uniform priors.
\end{abstract}


\section{Introduction}
\label{sec:intro}
Four pulsar timing array (PTA) collaborations recently published evidence for a correlated low-frequency signal thought to arise from a stochastic background of gravitational waves \citep[GWB][]{agazieNANOGrav15Yr2023b,reardonSearchIsotropicGravitationalwave2023,antoniadisSecondDataRelease2023a, xuSearchingNanoHertzStochastic2023}. 
Such a signal is expected to be generated by a population of supermassive black-hole binaries at the centers of galaxies \citep{agazieNANOGrav15Yr2023d,agazieNANOGrav15Yr2023c,antoniadisSecondDataRelease2023,antoniadisSecondDataRelease2024}, but more speculatively also by other sources \citep{afzalNANOGrav15Yr2023}. Bayesian inference lies at the heart of most of these analyses in the literature \citep{vanhaasterenMeasuringGravitationalwaveBackground2009,vanhaasterenNewAdvancesGaussianprocess2014}, and signal or noise parameters are typically quoted as Bayesian credible intervals. Even Frequentist detection statistics for gravitational waves \citep[GWs][]{allenHellingsDownsCorrelation2023a,gersbachSpatialSpectralCharacterization2024} usually make use of the noise models obtained by Bayesian inference \citep[][]{agazieNANOGrav15Yr2023b}.

Typically, the analysis of PTA data is carried out in two stages: firstly a noise analysis is run separately on each pulsar, and secondly a combined analysis is carried out on the combined all-pulsar dataset where models for correlated signals like GWs are included in the model. During the single-pulsar noise analysis data quality can be assessed, and noise models that are required for a joint search for GWs are tested in a variety of ways. The search for GWs in the all-pulsar dataset is computationally expensive, and much effort has gone into model pruning methods and computational tricks \citep[e.g.][]{ellisEfficientApproximationLikelihood2013,lentatiHyperefficientModelindependentBayesian2013,vanhaasterenAcceleratingPulsarTiming2013,lambRapidRefittingTechniques2023a,taylorParallelizedBayesianApproach2022,laalExploringCapabilitiesGibbs2023}.

Pruning the model of unnecessary components is an effective means to accelerate the analysis, but care must be taken in this step: a noise component that is not significantly detectable in the data may actually be present. Bayes Factors calculated during the single-pulsar noise analysis are often used to decide which components to prune.
In this letter we point out that this is an incorrect use of Bayes Factors, a statistical malpractice that falls in the category called ``circular analysis''~\citep{kriegeskorteCircularAnalysisSystems2009}.
The correct approach is to carry out model averaging, including both the model with-- \emph{and} the model without the noise components under consideration.
Model averaging has always been impractical due to its computational complexity.
We suggest that a computationally viable way of accomplishing model averaging exists in the form of Spike and Slab priors, which we prove to be identical to the already-used Log-Uniform priors in current PTA analysis. To summarize: PTAs need to include all possible noise sources with Log-Uniform priors which is mathematically identical to model averaging, or---where appropriate---make use of a Hierarchical Bayesian Model~\citep{vanhaasterenPulsarTimingArrays2024}.

\section{Model selection and model averaging}
\label{sec:circular_analysis}
Bayesian inference is based on the observation that the joint probability distribution of the data and model parameters can be divided into collection of conditional probabilities. Given data $d$, parameter(s) of interest $h$, nuissance parameter(s) $\theta$, and model hypothesis $M_{i}$ with $i \in [1,k]$ for $k$ different model hypotheses, we can write:
\begin{align}
    P(d,h,\theta,M_i) &= P(h,\theta,M_i | d) P(d) \\
                      &= P(d | h,\theta,M_i) P(h,\theta | M_i) P(M_i).
\end{align}
Here $P(d,h,\theta,M_i)$ is the joint distribution, $P(h,\theta,M_i | d)$ is the posterior distribution, $P(d)$ is the prior predictive (usually considered a normalization constant), $P(d | h,\theta,M_i)$ is the likelihood, $P(h,\theta | M_i)$ is the prior, and the $P(M_i)$ are the model prior odds. The various models $M_i$ usually represent different ways of modeling the signals of interest or the description of noise.
In the current paper, $M_i$ selects which noise components described by $\theta$ are included in the likelihood of the model.

When comparing how well different models fit the data, an often-calculated quantity is the fully marginalized likelihood (FML) or ``evidence'':
\begin{align}
    \label{eq:fml}
    P(d | M_i) &= \int {\rm d}h{\rm d}\theta \, P(d | h, \theta, M_{i}) P(h,\theta | M_{i}) \\
     &= P(M_{i} | d) P(d) / P(M_{i}).
\end{align}
The prior predictive $P(d)$ is usually unknown and left out\footnote{The prior predictive and FML are closely related and sometimes identical. Here they are only different because we consider multiple models $M_{i}$.}, and $P(M_i | d)$ can be normalized after the integral is carried out when $P(M_{i} | d)$ is needed. The FML of Equation~\eqref{eq:fml} is typically calculated using Nested Sampling, or the so-called odds ratio $\mathcal{O}_{ij}$ between model $M_{i}$ and $M_{j}$ is calculated using sophisticated Markov Chain Monte Carlo (MCMC) methods:
\begin{align}
    \label{eq:or}
    \mathcal{O}_{ij} &= \frac{P(M_i | d)}{P(M_j | d)} = \mathcal{B}_{ij} \frac{P(M_i)}{P(M_j)} \\
    \mathcal{B}_{ij} &= \frac{P(d | M_i)}{P(d | M_j)},
    \label{eq:bf}
\end{align}
where $\mathcal{B}_{ij}$ is called the Bayes Factor.

The main result of an analysis with Bayesian inference is often based on the posterior distribution of parameters of interest $h$, \emph{marginalized} (integrated) over the nuissance parameters and model space:
\begin{align}
    P(h | d) &= \sum_{i} \int {\rm d}\theta \, P(h,\theta,M_i | d) \\
    &\propto \sum_i \int {\rm d}\theta \, P(d|h,\theta,M_i)P(h,\theta|M_i)P(M_i).
\end{align}
This posterior distribution represents the combined marginalization over nuissance parameters and models, which is referred to as \emph{model averaging} in the Bayesian literature~\citep{gelmanBayesianDataAnalysis2013}. It is the appropriate way to summarize our ignorance, taking into account our prior beliefs while conditioning on the data.

\subsection{Circular analysis in pulsar timing}
\label{sec:incorrectpta}
From an experimentalist point of view it is considered incorrect to even look at the results of a (potential) analysis or even the data while deciding on the model. Indeed, in social sciences \citep{shadishExperimentalQuasiexperimentalDesigns2002} and controlled experimentation such as A/B testing \citep{kohaviTrustworthyOnlineControlled2020}, the model, expected results, and full methodology should be specified even before data is gathered.
In pulsar timing this is not the norm; it is routine to continuously look at the data and model summary statistics while manually working with and modifying the data. Historically and pragmatically this makes sense: the data is incredibly high signal-to-noise with respect to the timing model, and the timing model is so sophisticated that fully including all potential physical effects would be a daunting exercise, especially with the limited computational resources of past times. Moreover, since the data is so informative with respect to most timing model parameters, the process of model averaging would not change anything in the results.

The workflow under scrutiny in this paper is the ``single-pulsar noise analysis'' in pulsar timing array (PTA) projects, where data of individual pulsars are analyzed in order to understand the noise processes that are present in the data. Noise processes such as so-called ``intrinsic red noise'' (IRN) and more specific noise sources like band noise, are often not significantly detectable in the data of a single pulsar, and PTA projects typically calculate Bayes Factors to assess whether such processes are significantly present.
Using Bayes Factors as a criterion to decide whether noise processes are present in the data---and therefore should be included in the model---has been a common procedure since the advent of the Nested Sampling-based TempoNest \citep{ferozMultiNestEfficientRobust2009,skillingNestedSampling2004,lentatiTemponestBayesianApproach2014} and the analysis of the first data release of the International PTA (IPTA) \citep{lentatiSpinNoiseSystematics2016}.

In order to search for signals that induce correlated data between pulsars the entire set of all pulsars of a PTA needs to be analyzed simultaneously. The model that describes the combined data of all pulsars in full typically becomes computationally expensive to analyze, and in an attempt to limit the computational complexity the PTA projects strive to limit the number of model components that is included in the model. To achieve this, the above-mentioned Bayes Factors are frequently used to decide whether a model component needs to be included by setting a threshold on the Bayes Factor $\mathcal{B}_{ij} \ge \mathcal{B}_t$, where $\mathcal{B}_t$ is some threshold, such as $\mathcal{B}_t=5$ \citep{caballeroNoiseProperties422016,goncharovIdentifyingMitigatingNoise2021}. Additionally, so-called white noise parameters are determined and held fixed during the second/combined analysis.

Model selection has been an efficient and necessary means to prune the combined all-pulsars model, and therefore speed up the GW searches. However, upcoming computational improvements are making this model pruning less urgent. In that context, the calculation of Bayes Factors for noise characterization becomes largely unnecessary. For white noise parameters the practice of model pruning is acceptable, because those parameters are not covariant with the signals of interest.

Model selection for a subsequent analysis on the same data means that in Equation \eqref{eq:or} the prior odds $P(M_{i})$ are replaced by something else. Instead of having equal prior odds $P(M_{i})=P(M_{j})$ as assumed in standard single pulsar noise analysis, one sets all $P(M_{j})=0$ except for the one where $j\ne i$, where $P(M_{i}|d)$ had the largest FML during the single pulsar noise analysis. This is a circular use of data, where the results of an analysis are used to inform/tune the model for an analysis of the same data; a well-known practice in the statistics literature called ``circular analysis''~\citep{kriegeskorteCircularAnalysisSystems2009}.
Recent examples of this approach are seen in the analysis of the European PTA (EPTA) third data release \citep{antoniadisSecondDataRelease2023c} and the Chinese PTA (CPTA) first data release \citep{xuSearchingNanoHertzStochastic2023}, where Bayes Factors are used to deselect noise processes in their GWB analysis \citep{antoniadisSecondDataRelease2023e,antoniadisSecondDataRelease2023a}. The practice has been more widespread in the past several years, as the PTA community needed to prune the models for computational efficiency.

Using single pulsar noise analysis results to prune the model is not allowed because it effectively restricts the prior range based on the data. However, it is possible that it is discovered during the single pulsar runs that the model needs to be \emph{expanded} with extra model components that were initially left out. That is appropriate. An example is the extension of the pulsar binary model to include a Shapiro delay component, which are only present for certain orbit orientations.

\subsection{Interpretation of over circular analysis}
\label{sec:interpretation}
By pruning the model through the circular analysis practice of replacing $P(M_{i})$ in Equation \eqref{eq:or}, some pulsars that do not have significantly detectable noise processes do not have those noise processes modeled even if they may be present. The problem is that these signals could actually be there, evidenced by the fact that the PTA projects did have them in their prior when first analyzing their data during the single-pulsar noise analysis. If such signals are not significantly detected does not mean they are not potentially there.

In subsequent analysis of the whole array, it can turn out that these pulsars that do not have IRN or other noise components included in their model are contributing to the detection statistic of a GWB. Since the GWB in a single pulsar is also just a low-frequency signal, the signal that is actually present was not significant enough to show up in the Bayes Factor of the single pulsar analysis. Indeed, pooling together data can make weaker signals detectable that were originally not significant when looking at a single pulsar at a time.

By not including noise components like IRN for some pulsars in the model, one is effectively saying ``we know for a fact that this low-frequency signal is not noise, and we know for a fact that this low-frequency signal is gravitational waves''. The effect of circular analysis in PTAs is an overestimate of the signal (GWB) amplitude.

\subsection{Correct approach: model averaging}
\label{sec:spikeslab}
As described in Section \ref{sec:circular_analysis}, circular analysis as in the single-pulsar noise analysis using Bayes Factors is statistically incorrect. The correct analysis would make use of model averaging, correctly incorporating the covariance of signal and noise terms whether they are detectable or not. An often-heard fear among data analysts is that such a complicated/flexible model will reduce the sensitivity of the dataset significantly. While this fear is somewhat warranted, it is not a reason to not incorporate all appropriate model components. Not only will the data select for the correct model while down-weighting noise terms that are not there, it is also the correct model: the data analysts have actually searched for said noise terms.

In the literature, model averaging has been applied to pulsar timing using in various ways. General examples are Reversible-Jump Markov MCMC (RJMCMC) methods \citep{ellisTransdimensionalBayesianApproach2016}, which can be complex to implement efficiently. Product space sampling is more frequently used~\citep{arzoumanianNANOGrav11Year2018,taylorBrightBinariesBumpy2020} in the PTA community. In the appendix we prove that the usual Log-Uniform priors are equivalent to Spike and Slab priors if one appropriately re-interprets the prior bounds. Spike and Slab priors are, in turn, equivalent to model averaging~\citep{greenReversibleJumpMarkov1995}.
This makes model averaging even more straightforward than RJMCMC or product space sampling. In other words, instead of model selection, the PTA community needs to adopt a workflow where all potential noise sources are included with appropriate priors and so that model averaging is automatically performed. Fortunately, all PTA analyses are already set up to incorporate this workflow. Albeit somewhat slower, the analyses of PTA datasets actually becomes more straightforward when doing model averaging this way.
The general advice from the LISA global fit applies here: ``Model everything and let the data sort it out''~\citep{cornishBayeswaveBayesianInference2015}.

There are exceptions to the above. If noise components, like ``white noise'' processes, are reasonably expected to not be covariant with the signals of interest, then carrying out model selection for those processes is not going to affect the overall results. It is even perfectly acceptable to use the maximum a posteriori (MAP) estimates for those model parameters, and that significantly benefits computational efficiency.
The recommendation in this letter is to be conservative: apply model averaging to all model components that can be expected to be---even mildly---covariant with the signals of interest. The Log-Uniform priors need to be re-interpreted in terms of prior odds. When applicable, a Hierarchical Bayesian Model should be used instead of model averaging~\citep{vanhaasterenPulsarTimingArrays2024}. For IRN a Hierarchical model would likely be the most appropriate model.

\section{Example}
In order to demonstrate the effect of circular analysis, we provide an example where the incorrect procedure of circular analysis is followed. The mock data model is a simple analog of a contemporary pulsar timing array dataset, inspired by the MeerKAT PTA with $88$ pulsars.

\subsection{Toy model}
\label{sec:toy_model}
We use a model very similar to what was recently used in \citet[][section 2]{vanhaasterenPulsarTimingArrays2024}: we have data from $N_p=88$ pulsars each associated a random position on the sky, and each pulsar yields $N_o=6$ degrees of freedom from the lowest three frequency bins. This is roughly equivalent to the number of degrees of freedom that the MeerKAT Pulsar Timing Array (MPTA) is sensitive to in their current dataset \citep[][Matt Miles private communication]{milesMeerKATPulsarTiming2023}.

In our simplified model, there are only three processes that give a response in these pulsar data: the white noise process $X_{w}$, the IRN noise process $X_{r}$ and the signal process $X_{s}$.
Since we concentrate on a few low-frequency bins, we may assume that both $X_{w}$ and $X_{r}$ produce independent realizations, similar to in pulsar timing.
The process $X_{w}$ produces IID (independent identically distributed) data with a known amplitude in each pulsar, and $X_{r}$ produces an IID response in each pulsar with an unknown amplitude. Joined together, the total noise process $X_{n} = X_{w} + X_{r}$ yields:
\begin{equation}
    n_{ai} \sim \mathcal{N}(0, \sigma_{w,a}^{2} + \sigma_{r,a}^{2}),
    \label{eq:noise}
\end{equation}
where $n_{ai}$ is the $i$-th noise process sample at pulsar $a$, where we use the convention that $i$ labels observations $i \in [1,N_o]$ and $a$ labels pulsar $a \in [1,N_p]$. We denote a Gaussian distribution with mean $\mu$ and standard deviation $\sigma$ as $\mathcal{N}(\mu,\sigma^2)$. For pulsar $a$ the standard deviation of the noise is given by $\sigma_{a} = \sqrt({\sigma_{w,a}^{2} + \sigma_{r,a}^{2}})$, with $\sigma_{w,a}$ the white noise standard deviation, and $\sigma_{r,a}$ the unknown red noise standard deviation.

The signal process $X_{s}$ produces realizations that are correlated between pulsars. In a single pulsar, realizations of $X_{s}$ can still be modeled as IID and are therefore indistinguishable from $X_{r}$.
PTA projects focus mostly on Hellings \& Downs correlations, which represent the average correlations that an isotropically unpolarized ensemble of gravitational-wave sources would induce. These correlations, which we denote as $\Gamma(\zeta_{ab})$, depend only on the angular separation $\zeta_{ab}$ of pulsar $a$ and pulsar $b$:
\begin{align}
  \Gamma \left(\zeta_{ab}\right) &= \Gamma_{\rm{u}}\left(\zeta_{ab}\right) + \delta_{ab}\Gamma_{\rm{u}}(0) \nonumber \\
  \xi_{ab} &= \frac{\left(1 - \cos\zeta_{ab}\right)}{2} \nonumber  \\
  \Gamma_{\rm{u}}\left(\zeta_{ab}\right) &= \frac{3\xi_{ab} \log\xi_{ab}}{2} - \frac{1}{4}\xi_{ab} + \frac{1}{2}, 
\end{align}
where $\delta_{ab}$ is the Kronecker delta. The draws of $X_{s}$ in the data can now be written as:
\begin{align}
    s_{i} &\sim \mathcal{N}(0, \mathbf{\Sigma}) \nonumber \\
    \Sigma_{ab} &= h^2 \Gamma(\zeta_{ab}).
\end{align}
The parameter $h$ represents the signal amplitude, which has a similar interpretation as $\sigma_{a}$ for the noise, and $\mathbf{\Sigma}$ is the covariance of the multivariate Gaussian distributed variable $s_{i}$. The full data $d$ can then be written as:
\begin{equation}
    d_{ai} = s_{ai} + n_{ai},
    \label{eq:data}
\end{equation}
where we write $s_{ai}$ to mean the elements of $s$ of pulsar $a$ and observation $i$.

\subsection{Procedure and results}
\label{sec:results}
The data was generated using the following parameters: $\log_{10}\sigma_{w,a}^2 = 1.5$, $\log_{10}\sigma_{r,a}^2 = 1.5 \pm 0.2$ (Gaussian distributed around $1.5$, varied per pulsar), and $\log_{10}h^2 = 1.35$. The priors for signal $\log_{10} h^2$ and noise $\log_{10}\sigma_{r,a}$ were set to Log-Uniform priors between $10^{-3}$ and $10^{5}$. Pulsars were positioned uniformly across the sky.

The single-pulsar analysis for IRN is now essentially a one-dimensional problem: both the signal (GWB) and the noise (IRN and white noise) are identical. The white noise is known. Calculating Bayes Factors per Equation \eqref{eq:bf} is very quick. Model $M_{1}$---the ``Full'' model---includes all noise terms. Model $M_{2}$---the ``Data'' model---only includes $\sigma_{r,a}$ for a pulsar if the IRN Bayes Factor $\mathcal{B}_t \ge 5$. Additionally, in order to prevent posterior bias, we have also constructed a Hierarchical Bayesian Model (HBM) according to Equation (7) of \citet{vanhaasterenPulsarTimingArrays2024}, which we label as $M_{3}$.

\begin{figure}
    \includegraphics[width=0.48\textwidth]{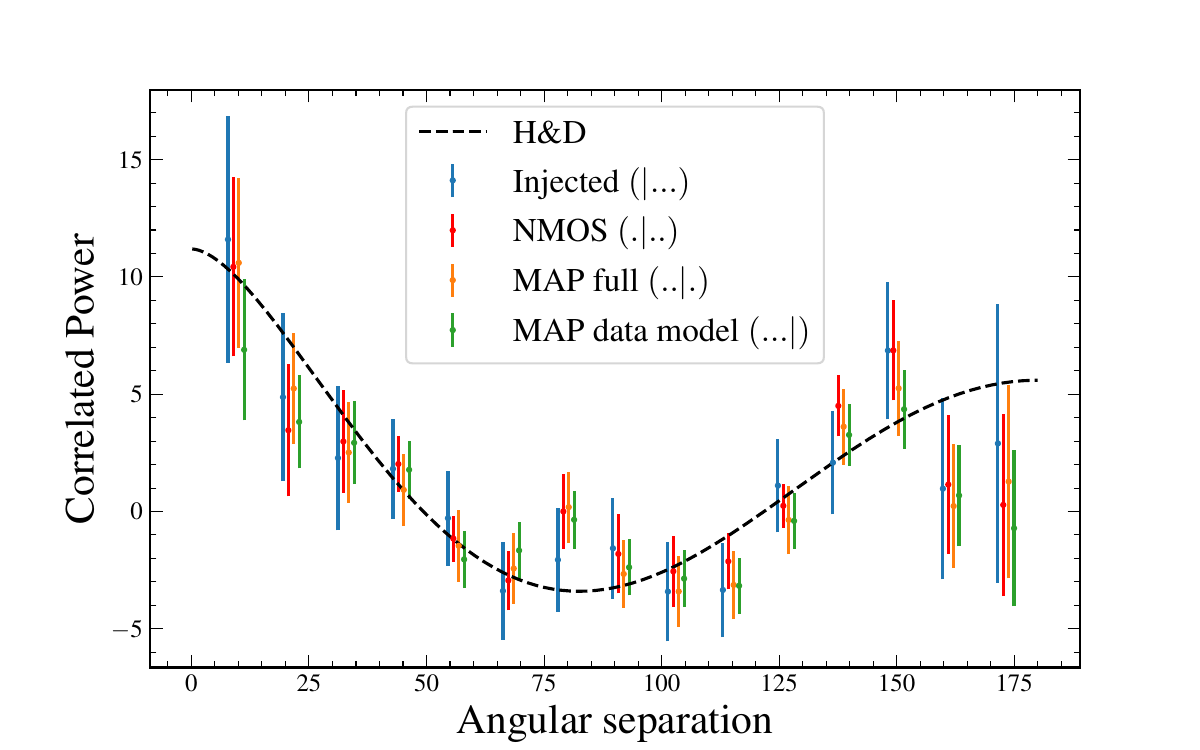}
    \caption{Cross power using the binned pair-covariant correlation statistic (PCOS) of \citet{allenHellingsDownsCorrelation2023a} for an array of pulsars. The PCOS requires all model parameters as fixed input parameters. For blue ``Injected'' (leftmost $(|...)$) the true injected parameter values were used. The red ``NMOS'' (second from left $(.|..)$) has been made with the Full model posterior samples. For orange ``MAP Full'' (second from right $(..|.)$) the joint Maximum A Posteriori values were used. For green ``MAP Data'' (rightmost $(...|)$), the circular analysis Bayesian MAP values were used, where for pulsars with insignificantly detected IRN the noise parameter values were set to zero. It is visible that the MAP Data model has smaller uncertainties than the other models.}
    \label{fig:pcos}
\end{figure}

\begin{figure}
    \includegraphics[width=0.48\textwidth]{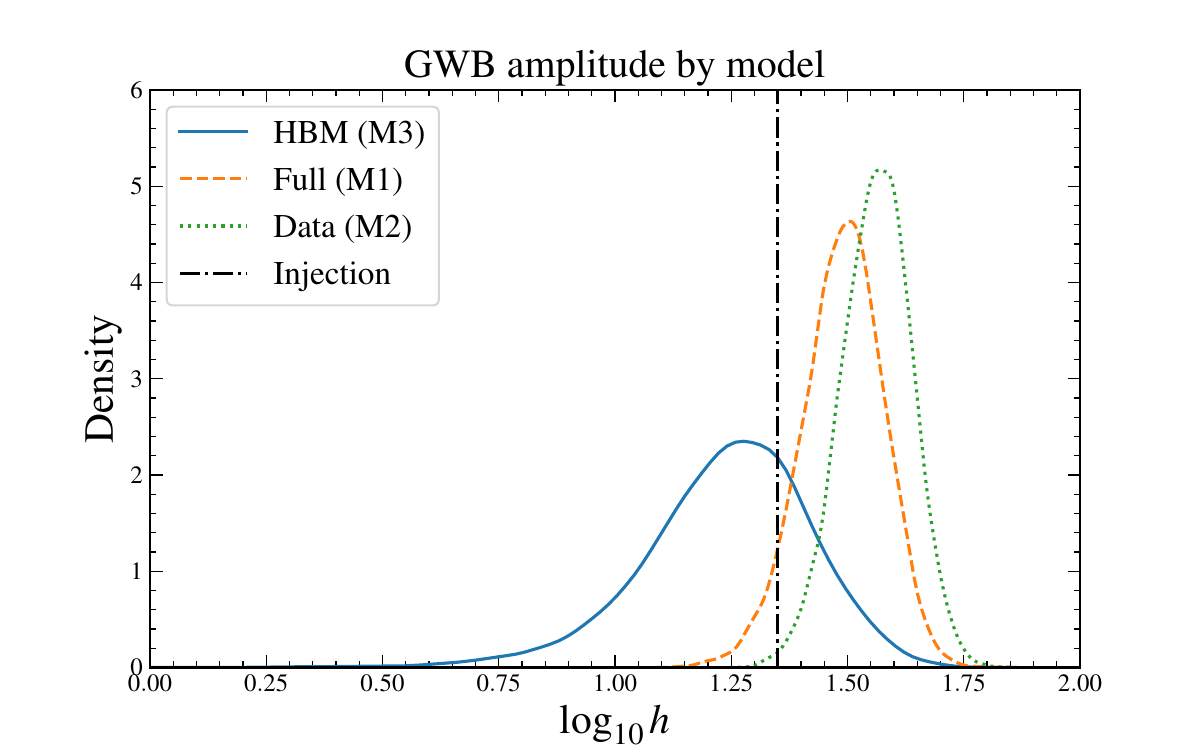}
    \caption{Posterior distribution of $\log_{10}h$ for the various models. The blue/solid ``HBM'' model represents a Hierarchical Bayesian Model (HBM) as in \citet{vanhaasterenPulsarTimingArrays2024}. The orange/dashed ``Full'' model includes all noise processes. The green/dotted ``Data'' model has the noise processes that did not have a Bayes Factor that surpassed the threshold excluded from the model. Although we expect the Full model to show posterior bias due to the lack of an HBM, we can see that the Data model showcases an extra over-estimate of $\log_{10} h$, which makes the posterior distribution incompatible with the injection value.}
    \label{fig:hcpost}
\end{figure}

The correlation deterction significance \citep{jenetDetectingStochasticGravitational2005} and the optimal correlation statistic \citep{anholmOptimalStrategiesGravitational2009,chamberlinTimedomainImplementationOptimal2015} can be used in a search for correlated stochastic signals in PTA data. In Figure \ref{fig:pcos} we use the pair-covariant optimal statistic (PCOS) introduced by \citet{allenHellingsDownsCorrelation2023a} combined with model parameter estimates from model $M_{1}$ and $M_{2}$. We also use the noise-marginalized optimal statistic (NMOS) adapted from \citep{vigelandNoisemarginalizedOptimalStatistic2018}, which is equal to the PCOS but now averaged weighted by the posterior distribution. In the figure,
we see that the circular analysis Data model has significantly smaller uncertainties than even the Full model. The Full optimal statistic does not incorporate covariance between $h$ and $\sigma_{a}$, so even the binned uncertainties of the Full model are underestimated compared to injections. Figure \ref{fig:hcpost} shows that the Data model displays more posterior bias than the Full model in a Bayesian analysis. As expected, the variance that was not modeled in some pulsars has now been attributed to the signal amplitude $\log_{10}h$, causing the estimate of $\log_{10}h$ to be biased high compared to the Full model. Also, as explained in \citet{vanhaasterenPulsarTimingArrays2024}, the HBM will exhibit less posterior bias than the Full model.

We stress that the example in this section is not meant to verify the validity of model averaging using Log-Uniform priors. The theoretical argument and proof in the appendix is sufficient. The toy model here is only a demonstration of what effect it has when one has a workflow of model selection: it has a demonstrable effect, and it necessarily introduces an artificial overestimate of the GWB amplitude. How large the effect is depends on the dataset. For some datasets the effect will be minimal, whereas for other datasets it might be substantial. We argue here that it is so easy to do model averaging or Hierarchical Bayesian Modeling that there is never a reason not to do it.

\section{Circular analysis in PTA literature}
Circular analysis is present in all contemporary PTA GW searches in the form of model pruning. For white noise parameters and other model components that are not covariant with the signals of interest this practice is acceptable. However, it leads to biased results if model pruning is applied to low-frequency signals that are covariant with the signals of interest.

In the most-recently released GW searches with PTAs, the Chinese PTA~\citep{xuSearchingNanoHertzStochastic2023} and the European PTA~\citep{antoniadisSecondDataRelease2023a} prune their models incorrectly using circular analysis on noise processes that are covariant with the GWB. The Parkes PTA (PPTA) does use Bayes Factors for model selection, but the important IRN and dispersion measure variations (DMV) processes that are covariant with the GWB are conservatively all included in the model~\citep{reardonSearchIsotropicGravitationalwave2023}, which makes the analysis correct. Moreover, the PPTA has done a very exhaustive search for noise processes that the other PTAs have not even attempted, making the analysis quite robust~\citep{reardonGravitationalwaveBackgroundNull2023}. NANOGrav includes fewer noise components than the PPTA, but IRN and DMVs are included so there are no concerns regarding circular analysis~\citep{agazieNANOGrav15Yr2023,agazieNANOGrav15Yr2023b}.

For all the above, we do not make any statements regarding how much effect circular analysis has on the actual numerical results. That is completely dependent on the dataset. All we are pointing out is that it is quite straightforward to do model averaging, and there is little reason in practice not to do it.

\section{Conclusions}
Bayes Factors are commonly used in PTA projects to select the noise sources that need to be included in the search for GWs as a means to prune the model and to gain computational efficiency. This workflow is statistically incorrect, and can lead to artificially increased GW amplitudes and detection significance reports \citep[][]{xuSearchingNanoHertzStochastic2023,antoniadisSecondDataRelease2023a}.
Instead, all model components for signal and noise need to be included in the full GW search in PTAs, with Log-Uniform priors on the amplitudes. We have included the proof that this is mathematically equivalent to model averaging over the whole set of possible models. Log-Uniform priors need to be re-interpreted in terms of the prior odds. Exceptions to this rule are signals that we have more prior information on, model components like white noise that are not covariant with signals of interest, and model components that should be modeled with Hierarchical Bayesian Models \citep{vanhaasterenPulsarTimingArrays2024}.

\section*{Acknowledgments}
I thank Boris Goncharov and my colleagues at NANOGrav for early comments on this letter. Jeremy Baier correctly pointed out that the practice under scrutiny should be referred to as circular analysis.

\section*{Data availability}
The data, code, and figures used in this article can be re-generated with a fully self-contained python script that is available online: \url{https://github.com/vhaasteren/circular_analysis}.

\appendix

\section{Proof of equivalence of Log-Uniform priors and Spike and Slab priors}
We assume that a signal or noise source contains an amplitude parameter $\theta = \log A$, where a Log-Uniform prior on $A$ means a flat prior on $\theta$ with prior bounds $\theta_{\rm min} \le \theta \le \theta_{\rm max}$. Using the same notation as in the rest of the paper, we are typically interested in some marginalized posterior, such as:
\begin{align}
    P(h|d) &= \sum_{i} \int {\rm d}\theta\, \frac{P(h,\theta,M_{i},d)}{P(d)} \\
    &= \sum_{i} \int {\rm d}\theta\, \frac{P(d|h,\theta,M_{i})P(h|M_{i})P(\theta|M_{i})P(M_{i})}{P(d)}.
    \label{eq:ma}
\end{align}
To simplify things, we assume we only have two models: $M_{1}$ indicating a signal with amplitude parameter $\theta$ is present in the model, and $M_{2}$ indicating that signal is not present in the model. The prior $P(
\theta|M_{1})$ is constant for $\theta_{\rm min} \le \theta \le \theta_{\rm max}$ and zero elsewhere. Model $M_{2}$ does not depend on $\theta$. We also assume $P(h|M_{1})=P(h|M_{2})=P(h)$.
Moreover, the $M_{1}$ prior range on $\theta$ is such that $\theta_{\rm min}$ is so small that it is indistinguishable from not having the signal in the model. In other words, if $\theta_{\rm low} \le \theta_{\rm min}$, then:
\begin{equation}
    P(d|\theta=-\infty,h,M_{1}) = P(d|\theta=\theta_{\rm low},h,M_{1}) = P(d|h,M_{2}).
\end{equation}
We are therefore allowed to re-define model $M_{2}$: model $M_{2}$ now has the same likelihood as $M_{1}$ which depends on $\theta$, and we give it a prior constant in $\theta$ with different bounds. As long as $\theta \le \theta_{\rm min}$, the likelihood of model $M_{2}$ is identical to that of model $M_{1}$. Thus, we redefine the $M_{2}$ prior:
\begin{equation}
    P(\theta|M_{2}) = 
    \begin{cases} 
    \frac{1}{\theta_{\rm min} - \theta_{\rm min2}} & \text{for } \theta_{\rm min2} \leq \theta \leq \theta_{\rm min} \\
    0 & \text{otherwise}
    \end{cases},
    \label{eq:nowprior2}
\end{equation}
where we define a lower-limit $\theta_{\rm min2}$ which we are free to choose as long as $\theta_{\rm min2} \le \theta_{\rm min}$. This turns Equation \eqref{eq:ma} into
\begin{align}
    \label{eq:sspr}
    P(h|d)P(d) = \int {\rm d} \theta\, P(h) &\Big[ P(M_{1})P(\theta | M_{1})P(d | \theta, h, M_{1}) +  \nonumber \\
    & P(M_{2})P(\theta | M_{2})P(d | \theta, h, M_{2}) \Big] \\
     = \int {\rm d} \theta\, P(h) &P(M_{0})P(\theta | M_{0})P(d | \theta, h, M_{0}),
\end{align}
where we have introduced model $M_{0}$ in the last line, and $P(M_{0})=P(M_{1})+P(M_{2})=1$. The model $M_{0}$ specification is identical to that of model $M_{1}$: the same likelihood definition, and a prior flat in $\theta$. The only difference is that the prior range on $\theta$ in model $M_{0}$ is now set to $\theta_{\rm min2} \le \theta \le \theta_{\rm max}$, and:
\begin{equation}
    \theta_{\rm min2} = \theta_{\rm min} - \frac{P(M_{2})}{P(M_{1})} (\theta_{\rm max} - \theta_{\rm min}).
\end{equation}
In other words, the Log-Uniform prior model $M_{1}$ is identical to the Spike and Slab \citep{greenReversibleJumpMarkov1995} prior of model $M_{0}$, provided we interpret the prior bounds on $\theta$ appropriately. Log-Uniform priors can do the same thing as trans-dimensional analysis \citep{ellisTransdimensionalBayesianApproach2016} for model averaging.

Another useful way to think about the above derivation makes the connection with the Savage-Dickey density ratio \citep{dickeyWeightedLikelihoodRatio1971}, which is the ratio of the average likelihood over the prior and the likelihood at $A=0$. In general, the FML is the average likelihood taken over the prior, meaning that Bayes Factors are ratios of average likelihoods.
In the context above, $P(d|\theta=\theta_{\rm low},h,M_{1})$ does not vary over the prior $\theta_{\rm min2}\le \theta\le \theta_{\rm min}$. It trivially follows that the Bayes Factor between $M_{1}$ and $M_{2}$ is identical to the Savage-Dickey density ratio between model $M_{1}$ and model $M_{2}$:
\begin{equation}
    \mathcal{B}_{12} = \frac{\int {\rm d}\theta\, P(d|\theta,h,M_{1})P(\theta | M_{1}) } {P(d|h,M_{2})}.
\end{equation}

\bibliographystyle{mnras}
\bibliography{references}

\end{document}